# Legal Document Retrieval using Document Vector Embeddings and Deep Learning

Keet Sugathadasa*, Buddhi Ayesha*, Nisansa de Silva*, Amal Shehan Perera*,
Vindula Jayawardana*, Dimuthu Lakmal*, Madhavi Perera[†]
*Department of Computer Science & Engineering
University of Moratuwa
[†]University of London International Programmes
University of London
Email: keetmalin.13@cse.mrt.ac.lk

*Abstract*—Domain specific information retrieval process has been a prominent and ongoing research in the field of natural language processing. Many researchers have incorporated different techniques to overcome the technical and domain specificity and provide a mature model for various domains of interest. The main bottleneck in these studies is the heavy coupling of domain experts, that makes the entire process to be time consuming and cumbersome. In this study, we have developed three novel models which are compared against a golden standard generated via the on line repositories provided, specifically for the legal domain. The three different models incorporated vector space representations of the legal domain, where document vector generation was done in two different mechanisms and as an ensemble of the above two. This study contains the research being carried out in the process of representing legal case documents into different vector spaces, whilst incorporating semantic word measures and natural language processing techniques. The ensemble model built in this study, shows a significantly higher accuracy level, which indeed proves the need for incorporation of domain specific semantic similarity measures into the information retrieval process. This study also shows, the impact of varying distribution of the word similarity measures, against varying document vector dimensions, which can lead to improvements in the process of legal information retrieval.

## I. INTRODUCTION

Similarity measures between words, sentences, paragraphs, and documents are a prominent building block in majority of tasks in the field of Natural Language Processing. Information Retrieval (IR) is an application that heavily uses similarity measures. This is due to the fact that the function of IR algorithms require the identification of similarity and context of interest. Thus when the IR is done on textual documents, semantic similarity measures play a major role in both identifying documents related to the query and ranking the resultant documents according to the relevance [1].

Even though document retrieval is a well researched and mature area of study, it becomes a different and unique problem for each domain of research. This is mainly due to the syntactical complexities in the textual documents and the semantic structure of the text. Textpresso [2] is a text mining, information extraction, and information retrieval system that goes far beyond traditional keyword search engines, built for biological literature. Further in the biological domain, OmniSearch [3] is a semantic search system based on the ontology for microRNA-target gene interaction data [4]. All other fields such as medicine [5], geology [6], and music [7] also have their own unique aspects which make the information retrieval task more complex and domain specific.

In terms of carrying out a legal case in particular and also as the courts are binding upon precedent to know the law and knowing under which case that the laws were established and enforced is of utmost importance. There are also instances of which courts turn into case law of other countries in the absence of their own that explains the case context at hand. Amongst the large number of cases and the phases of which the cases have evolved, it remains impossible for those in the field of law to remember and have cases and laws in memory.

Because of these reasons, we selected the legal document retrieval as our domain. The legal domain, which is our primary focus in this study, contains considerable amount of domain specific jargon where the etymology lies with mainly Latin and English, which makes the information retrieval (IR) task multilingual. To complicate this fact, in certain cases, the meaning of the words and context differs by the legal officers' interpretations. This is the main standout for current legal IR systems such as Westlaw[1] and LexisNexis[2]. The main drawback of these systems is that, they still being boolean indexed systems, extensive user training is required of a legal professional to utilize them [8]. Despite this shortfall, Westlaw and LexisNexis have the largest number of paying subscribers for legal information retrieval, which indeed validates the need for a legal information retrieval system.

This study targets the information retrieval for the legal domain where experiments are being carried out over 2500 legal cases collected from Findlaw [9] and other online legal resources via both retrieval and extraction. We propose a system that includes a page ranking graph network with TF-IDF to build document embeddings by creating a vector space for the legal domain, which can be trained using a neural network model supporting incremental and extensive training for scalability of the system.

The structure of the paper is as follows. Section II gives a brief overview on the current tools being used and domains that have tackled this problem of domain specific information retrieval. Section III gives a description on the methodology being used in this research in order to obtain the results and conclusions as necessary. That is followed by Section IV that presents and analyses results. The paper ends with Section V which gives the conclusion and discussion on future work.

---
[1]https://www.westlaw.com/
[2]https://www.lexisnexis.com/

## II. Background and Related Work

This section illustrates the background of the techniques used in this study and work carried out by others in various areas relevant to this research. The subsections given below are the important key areas in this study.

### A. Information Retrieval

Information retrieval is finding material of an unstructured nature that satisfies an information need from within large collections of documents [8]. The task of document retrieval has far more reach into research areas [10] such as video/song identification [11], newspaper categorization and retrieval [12], and multilingual document retrieval systems. Most of these systems commonly use skip-grams, Bag of Words (BOW), or term frequency inverse document frequency (TF-IDF) [13]. Day by day, the field of information retrieval get optimized to provide better results for users' information needs.

The modest way of identifying similarities between documents is by measuring the distance between each of the documents in a given vector space. But the problem arises with the way we represent documents in order to get a more accurate and precise representation for each of the documents. As mentioned in this study, previous studies have also tried to address this problem with various combinations of weighting mechanisms. Salton has addressed this issue by trying different combinations of statistic measures and term frequencies [10], whereas Perina has come up with a Componential Counting Grid for learning document representations [14].

TF-IDF [13] is a text mining technique, that gives a numeric statistic as to how important a word is to a document, in a collection or a corpus. This relative importance of words in a document is used to retrieve and categorize documents in relation to one or more query words. According to our research, the *term frequency* is calculated using equation 1, where $T_d$ is the most occurring term in document $d$. $TF_{t,d}$ is the frequency of term $t$ in document $d$ and the *inverse document frequency* is calculated using the equation 2, where $IDF_{t,D}$ is the inverse document frequency of $t$ in a corpus $D$. Finally, the $TF\text{--}IDF_{t,d,D}$ value is calculated using equation 3.

$$TF_{t,d} = 0.5 \; + \; 0.5 \, (\frac{term \; t \; count \; in \; d}{count \; of \; term \; T_d}) \quad (1)$$

$$IDF_{t,D} = \log{(\frac{total \; number \; of \; documents \; in \; D}{number \; of \; documents \; with \; t \; in \; it})} \quad (2)$$

$$TF\text{--}IDF_{t,d,D} = TF_{t,d} \times IDF_{t,D} \quad (3)$$

### B. Lexical Semantic Similarity Measures

Applications of text similarity measures include, relevance feedback classification [15], automatic text summarization by text restructuring [16], automatic evaluation of machine translation [17], semi-supervised ontology population [18], and determining text coherence [19]. Some approaches are based on statistical methods [20], vector representations, string/corpus based approaches, and hybrid similarity measures where four similarity measures were tested in [21] and eight similarity measures were tested in [22].

TF-IDF alone uses an inverse document frequency for term frequencies where it does not consider surrounding context amongst words in a text. This mapping is simply done using count and probabilistic measures.

### C. Vector Similarity Measures

In the field of information retrieval, the similarity between documents or terms are measured by mapping them into a vector of word frequencies in a vector space and computing the angle between the pair of vectors [23]. Distance and similarity measures encounter in various fields like chemistry, ecology, biological taxonomy and so on. Some of the traditional approaches used for similarity/dissimilarity measures include Hamming Distance [24, 25], inner product, Tanimoto distance etc. Sung-Hyuk Cha [26] addresses similarity measures available in terms of both semantic and syntactic relationships which are being used in various information retrieval problems. Sung-Hyuk Cha et al. [27] also describes how vector similarities can be obtained by enhancing binary features on vectors. Jayawardana et al. has utilized vector similarity measures in deriving representative vectors for ontology classes [28].

The word similarities are calculated according to the distance in space between two vectors. Wu and Palmer proposed a method to give the similarity between two words in the 0 to 1 range [29], where they have further used the cosine distance [30] as a measure of similarity.

The cosine similarity is measured by the angle between the two vectors. If they are parallel, the cosine distance is equal to one and the vectors are said to be equal. If the vectors are perpendicular, they are said to be dissimilar (no relation to each other). This similarity is based on terms or words in the legal domain. For example, it will show that words like father, mother, family are falling as similar words whereas tree, book and king falling as dissimilar words.

Similarly, the relationship between vectors can be measured in terms of the displacement between the two vector points. This is really helpful when the relationship between two elements is not known. For example, let's assume that we know the displacement between the vectors of man and woman, and the vector direction. Then if we need to find a similar relation for king, the model will find a vector in the same direction with a similar displacement from the point of the king vector, and will return queen as the answer.

The similarity and relationships between legal terms are very complex, where some relationships are based on hierarchical models. By the vector space obtained via word2vec, our study can prove many similarities and relationships which are not possible in a general word2vec implementation [31]. With this, we intend to show that results tend to be more accurate when the entire process is integrated with NLP techniques as described in this study.

*D. Legal Information Systems*

One of the leading domains, that severely adhere to this issue is the medical profession. The Canon group has come up with a representation language [32] for medical concepts in 1994, whereas word representations for hospital discharge summaries [33] was developed in 2013.

Schweighofer [34] claims that there is a huge vacuum that should be addressed in eradicating the information crisis that the applications in the field of law suffer from. This vacuum is evident by the fact that, despite being important, there is a scarcity of legal information systems. Even though the two main commercial systems; WestLaw[3] and LexisNexis[4] are widely used, they only provide query based searching, where legal officers need to remember keywords which are predefined when querying for relevant legal information. Hence, there is still a hassle in accessing this information.

One of the most popular legal information retrieval systems is KONTERM [34], which was developed to represent document structures and contents. However, it too suffered from scalability issues. The currently existing implementation that is closest to our proposed model is Gov2Vec [35], which is a system that creates vector representations of words in the legal domain, by creating a vocabulary from across all corpora on supreme court opinions, presidential actions, and official summaries of congressional bills. It uses a neural network [36] to predict the target word with the mean of its context words' vectors. However, the text copora used here, itself was not sufficient enough to represent the entire legal domain. In addition to that, the Gov2Vec trained model is not available to be used by legal professionals or to be tested against.

There are certain semantic and syntactic relationships, which are very specific for a domain like law, as it is to other domains like medicine and astronomy. Even though they point out that these relationships can be included in the training process, it seems like there is a lot more work to be done due to the special syntactic and semantic behavior in the legal domain.

Languages being used are sometimes mixed up with several origins (i.e English, Latin etc) and in certain cases, the meaning of the words and context differs by the legal officers' interpretations. One of the popular systems named KONTERM [34], is an intelligent information retrieval system that was developed to represent document structures and contents, to address this issue.

This is where Bag Of Words techniques, along with sentence similarities, play a major role in not only coming up with a representation scheme for words and phrases using a vector space, but also in identifying semantic similarities between the documents, which can be used in many applications for context identification and other purposes.

---

[3]https://www.westlaw.com/
[4]https://www.lexisnexis.com/

*E. TextRank Algorithm*

*TextRank* algorithm [37], is based on Google's *PageRank* algorithm [38]. *PageRank* (PR) was used by Google Search to rank websites in their search engine results. *TextRank* uses *PageRank* to score the sentences in a document with respect to other sentences in the document by using the graph representation $G$, for the set of sentences $S$, where the edges in set $E$ represent the similarity between each of these sentences in the given document.

$$G = \{S, E\} \qquad (4)$$

The first step is to separate the document into sentences to be used by TextRank. Next, a bag of words for each of the sentences is created. This is an unordered collection of word counts relevant to each sentence. This creates a sparse matrix of words with counts for each of them. Then a normalization technique is used for each count, based on TF-IDF. The next step is to create a similarity matrix $A$ between sentences, which shows the relevance and similarity from one sentence to another. It is then used to create a graph as shown in Fig.1. A threshold is used to decide which edges would stay in the final graph.

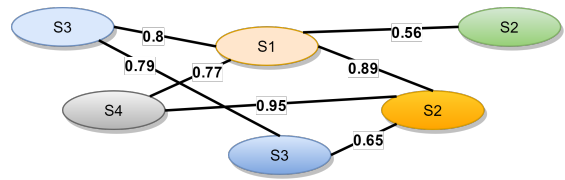

Fig. 1. Sentence Similarity Graph Network

Finally the PageRank algorithm will be used here to score the edges of the graph.

## III. METHODOLOGY

This section describes the research that was carried out for this study. Each section below, addresses a component in the overall methodology. Initially, we collected over 2500 legal case documents for this study from the FindLaw [9] website using multi-threaded webcrawlers. Hereafter in this study, we refer to this legal case document collection as the text corpus. An overview of the methodology we propose in this system is illustrated in Fig.2.

The following sub sections describe the steps that were carried out in this study, where we generate three unique document vector models: one is generated with raw document references, another using a neural network, and the other one is generated using sentence ranking derived from the domain specific semantic similarity measures, presented in [21]. Each document vector model is used to train a neural network with n-fold cross validation, which is then tested against a golden standard benchmark to generate the final results.

*A. Document Relevance Mention Map*

For this study, we defined a single legal case as a document unit, which gives us a granularity level [39] to contain many legal case documents pertaining to different legal areas. Given that, legal cases are reports of the proceedings of a lawsuit. It is possible to observe that prior legal cases are mentioned in them as case law references. By definition, these are the legal cases that the lawyers involved in the case, deemed relevant to the case at hand. Given that this study is also concerned with finding relevant legal cases for a selected legal case, it is imperative that we use the mentioned legal cases for the training of the models. Hence, we defined the *inverted index* structure, where for each document, we maintained a list of documents, which were referenced in that text. A list of this nature is called the *postings list*, where each element in the list is defined as a *posting* [8]. We use a *Dictionary* data structure to store the document vocabulary and a linked list for each of the posting lists. This overall structure is illustrated in Fig. 3. The size of the document vocabulary ($n$), is the number of document units contained in the text corpus.

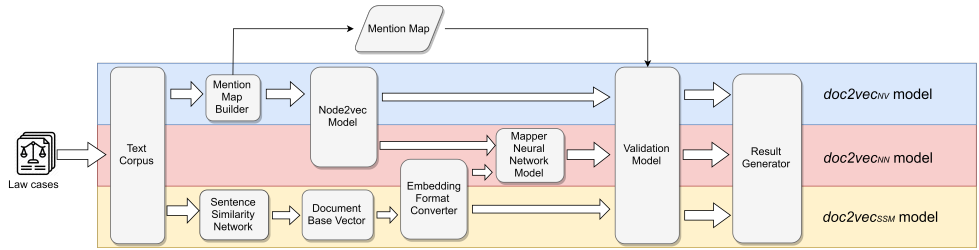

Fig. 2. Flow diagram for the Overall Methodology

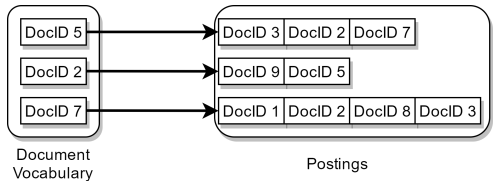

Fig. 3. Document Relevance Mention Map

The listing of mentions and references of other legal cases was a non-trivial Information Extraction (IE) task as discussed by [8]. This was mainly due to the difference in abbreviations and legal jargon that were used in online repositories. The first problem was the lack of a standard and proper naming convention for the legal case names. For example, depending on the case, the word *corporation* was shortened as *corp* or *co.* even when referring to the same legal case title, despite what is on record as the proper legal case name. This problem was solved by regular expression-based Information Extraction [2]. The second problem was the legal name reference using abbreviations of the case name in the text body. For example *Office of Workers' Compensation Programs* was referred as *OWCP*, in the body text of the cases that referred cases pertaining that entity despite the fact that all the cases that were being referred were using the full name of the entity in their titles. To solve this problem, extensive searching on the Findlaw Search Engine was done, and mention names were matched with the legal case document names.

This mapping from each document, to other documents which refer to the original document, is called the *mention map* ($M$). Equation 5 shows the formal definition of $M$, which contains $n$ number of keys. Each key $m$, has an associated postings list $l$, that contains the document IDs which have referenced in $m$.

$$M = \begin{bmatrix} m_1 & m_2 & m_3 & \dots & m_n \\ l_1 & l_2 & l_3 & \dots & l_n \end{bmatrix} \quad (5)$$

This is the base component of our study, which will be directly used as input to train the $doc2vec_{NV}$ model and to validate the neural network as shown in Section III-G. For the $doc2vec_{SSM}$ model, we carried out a number of advanced computational steps which will be discussed in the following sections. Incorporating these two models into another unique model, using a neural network, is given in the $doc2vec_{NN}$ model. In Section IV, we compare and contrast these three models and show the improvement of accuracy that can be gained by the following domain specific enhancements.

*B. Sentence Similarity Graph Network*

In a text document, sentences play a vital role to its semantic structure [40]. This is mainly due to the fact that the semantic information contained in the words in sentences are an accurate representation of the overall semantics of the document. Thus, in the attempts to represent a document as a feature vector, one of the most common methods is the *bag of words* [41] approach, which yields acceptable results even-though each word is considered independent of the surrounding words in the context. In this study, we created a sentence similarity graph network using the semantic similarity measures between sentences by the *TextRank* algorithm [37] described in Section II-E. We used a threshold of $0.5$ to determine the level of similarity between nodes in the graph.

We created a document corpus, which is a subset of the entire text corpus, where we picked the most important sentences in each of the documents, from the sentence similarity graph network. We selected the $k$ most important sentences within a document, using the sentence similarity graph network generated, with $|S| \geq k$, where $S$ is the set of sentences within a document. Hereinafter, this subset will be referred to as the *document corpus*.

*C. Text Preprocessing*

First, we pre-processed the *document corpus* to clean the text of unwanted characters and common words, in order to obtain the optimal size for the final vocabulary $V$. The pipeline that we used in this study is illustrated in Fig.4. In the linguistic preprocessing step, we used lemmatizing and case-folding to lowercase, as our primary Natural Language Processing(NLP) techniques. We used the Stanford Core NLP [42] library for this purpose.

*D. Document Base Vector Creation*

We ran TF-IDF [13] on the *document corpus* and built a TF-IDF weight matrix $T$ between the terms and documents. $T$ is a $v \times n$ matrix, where $v$ is the size of vocabulary vector $V$. Matrix $T$ is shown in Equation 6, where $t_{i,j}$ is the TF–IDF value of term $i$ in document $j$ of corpus $D$ calculated according to Equation 3.

$$T = \begin{bmatrix} t_{1,1} & t_{1,2} & t_{1,3} & \ldots & t_{1,n} \\ t_{2,1} & t_{2,2} & t_{2,3} & \ldots & t_{2,n} \\ \ldots & \ldots & \ldots & \ldots & \ldots \\ t_{v,1} & t_{v,2} & t_{v,3} & \ldots & t_{v,n} \end{bmatrix} \quad (6)$$

Next we defined GTF (Global Term Frequency) as shown in Equation 7. GTF is an extension of Equation 1 that was used to calculate TF. Here, $n$ is the number of legal case documents considered.

$$GTF_{t,D} = \frac{\sum_{i=1}^{n} TF_{t,i}}{n} \quad (7)$$

Then we scaled both the GTF and IDF values as according to Scalings 8 and 9.

$$0 \leqslant GTF \leqslant 1 \quad (8)$$

$$0 \leqslant IDF \leqslant 1 \quad (9)$$

Next we pruned the GTF and IDF values as given in Equations 11 and 13, where $\alpha$ values are given in Equation 14. After scaling the GTF and IDF values obtained from above, we calculate the mean separately for both of them, as $\mu_{GTF}$ and $\mu_{IDF}$. Next step is to calculate the standard deviation separately for both of them, as $\sigma_{GTF}$ and $\sigma_{IDF}$. The $\alpha$ value represents the factor by which the standard deviation is multiplied, and the range of the dataset is selected. This is depicted in Fig.5.

$$r_{GTF} = \alpha \times \sigma_{GTF} \quad (10)$$

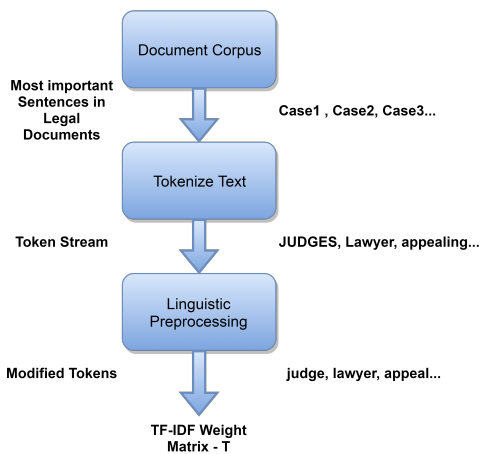

Fig. 4. Preprocessing Pipeline

$$GTF_{t,D} = \begin{cases} x_{t,D} & \text{if } \mu_{GTF} - r_{GTF} \leqslant x_{t,D} \leqslant \mu_{GTF} + r_{GTF} \\ 0, & \text{otherwise} \end{cases} \quad (11)$$

$$r_{IDF} = \alpha \times \sigma_{IDF} \quad (12)$$

$$IDF_{t,D} = \begin{cases} x_{t,D} & \text{if } \mu_{IDF} - r_{IDF} \leqslant x_{t,D} \leqslant \mu_{IDF} + r_{IDF} \\ 0, & \text{otherwise} \end{cases} \quad (13)$$

$$\alpha = [0.1, 0.5, 1, 2, 3] \quad (14)$$

Using GTF, we defined GTF-IDF in Equation 15 as an extension to TD-IDF given in Equation 3, to generate a $GTF{-}IDF_{t,D}$ value for each term in the vocabulary $V$. In GTF-IDF, the IDF algorithm is the same as Equation 2. The objective of GTF-IDF is to identify the words that are most significant to the domain $D$.

$$GTF{-}IDF_{t,D} = GTF_{t,D} \times IDF_{t,D} \quad (15)$$

We defined $V'$ by sorting $V$ by the descending order of $GTF{-}IDF_{t,D}$ values and obtained the common base vector template ($B$) as shown in Equation 16. $B$ is a word vector of length $p$, such that the $p$ number of words that are most significant to the domain $D$ are contained in $B$. Thus, $B$ was obtained by taking the first $p$ elements of the sorted $V'$.

$$B = \{term_1, term_2, ..., term_p\} \quad (16)$$

*E. Sentence Similarity Based Document Vector Representation*

Next, we created a vector representation for documents in the text corpus. As mentioned in Section III-D, the vector representation of a document was a $p$ dimensional vector, representing the most important $p$ terms from the context on the entire text corpus, as shown in Equation 16. As mentioned in Section III-B, we selected the $k$ most important sentences within a document, with $|S| \geq k$, where $S$ is the set of sentences within a document. For the $j$th document, we defined $R_j$ to be the total set of unique words in the selected $k$ sentences. We defined the $Seek$ function as shown in Equation 17, which would return the index $i$ of word $w$ given a vocabulary of words $U$.

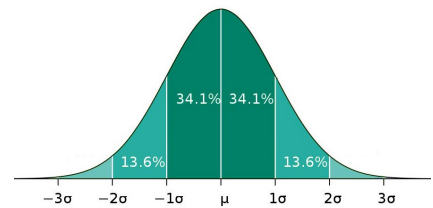

Fig. 5. Distribution of GTF and IDF values with a factor of standard deviation

$$i = Seek(w, U) \tag{17}$$

Finally, we defined the document vector representation $H_j$ for the $j$th document as shown in Equation 18, by calculating each $h_{j,i}$ as given in Equation 19, where $b_i$ is the $i$th element of the document vector template $B$ created by Equation 16, and $t_{a,j}$ is the element at row $a$ column $j$ of matrix $T$ defined in Equation 6. Each document vector $H_j$ was normalized using the L2 Normalization [43].

$$H_j = \{h_{j,1}, h_{j,2}, ..., h_{j,i}, ..., h_{j,p}\} \tag{18}$$

$$h_{j,i} = \begin{cases} t_{a,j} & \text{if } b_i \in R_j, a = Seek(b_i, V) \\ 0, & \text{otherwise} \end{cases} \tag{19}$$

*F. Scalable Feature Learning Node2vec Model*

As depicted in Fig.2, this study built two unique models: $doc2vec_{NV}$ model and the $doc2vec_{SSM}$ model. The $doc2vec_{NV}$ model was generated using an algorithm known as Node2vec [44], which is a wrap around the word2vec model introduced by Thomas Mikolov [31]. The node2vec algorithm is a scalable feature learning technique for networks, which learns continuous representations for nodes in any (un)directed or (un)weighted graph.

The input for this algorithm was an Edgelist, which contained the list of edges generated from the mention map in Section 3. Each document in the document vocabulary was paired with each of their references separately, and that was used to generate the list of edges. If *DocID 2* had referenced *DocID 9* and *DocID 5*, then the Edgelist would be pairs as *DocID 2, DocID 9*, and *DocID 2, DocID 5*

Finally, the output was a vector space that contains a set of feature vectors for the list of documents in the legal text corpus.

*G. Mapper Neural Network Model*

The Mapper Neural Network model $doc2vec_{NN}$, was trained using the both models: $doc2vec_{NV}$ model and the $doc2vec_{SSM}$ model. These two models were trained separately on different vector spaces, but with the same set of document IDs. Therefore, it is mandatory to build a model that could incorporate the different features in both of these vector spaces and produce the document ID of a given legal document, when the corresponding document ID in the other vector field, is provided. This process is depicted in Fig.6

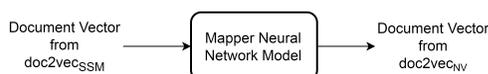

Fig. 6. Mapper Neural Network Input and Output

*H. Experiments*

As mentioned in Section III-G, we generated the $doc2vec_{NV}$ model and the $doc2vec_{SSM}$ model separately, which were later used to obtained the $doc2vec_{NN}$ model. In this study, our experiments are to compare and contrast the accuracy levels of these three models to varying $p$ values, where $p$ is the dimension of the document base vector $B$, as given in Equation 16. The $p$ values used in this study were; 250, 500, 750, 1000 and 2000. The accuracy measures we use for this study are based on recall [45], of which the formula is given by Equation 20, where $\{Relevant\}$ is the set of relevant documents taken from the golden standard and $\{Retrieved\}$ is the set of documents requested from the model. $\{Relevant\}$ is the number of unique references each *DocID* has got in the document mention map, which is depicted in Fig.3 in section 3. In other words, this is the length of $l$ (posting list) of each document, according to Equation 5.

$$Recall = \frac{\mid \{Relevant\} \cap \{Retrieved\} \mid}{\mid \{Relevant\} \mid} \tag{20}$$

For each of the different models, we used ten-fold cross validation [46] to measure the accuracy levels. The final results were further validated with the help of legal domain experts, to ensure that the results were accurate as expected.

## IV. RESULTS

The results obtained in this study is given in Table I, with varying $\alpha$ and $p$ values. Fig.7(a) and Fig.7(b) illustrate variation in recall value corresponding to varying $\alpha$ and $p$ values. Fig.7(a) depicts the recall variation in $doc2vec_{SSM}$ model where Fig.7(b) depicts the same in $doc2vec_{NN}$ model. This section of the study contains results obtained from the three models, where the recall values are calculated based on the golden standard measure, as mentioned in Section III-H.

The recall values obtained from the $doc2vec_{SSM}$ model, shows a slight degradation compared to the other models, where it involves a set of semantic similarity measures, in order to get the final result, as given in Section III. On the other hand, the $doc2vec_{NV}$ model shows significant improvements in results, in terms of recall, where it was generated using the Mention Map, from Section III-A.

The $doc2vec_{NN}$ model, which has a comparatively higher level of recall values compared to both the descendant models, is an ensemble model generated by both the above models, which indeed proves the research objectives of this study, which was to prove that domain specific document representation and retrieval models, need domain specific techniques to cater to the given information need. This is by incorporating domain specificity into this ensemble model.

The Fig.7(b) shows a 3-dimensional plot of the document vector representation accuracies corresponding to ensemble $doc2vec_{NN}$ model, obtained from our research study. These values were taken against, different $p$ values with varying $\alpha$ values. As shown, higher the $p$ value, higher the accuracy of the document vector implementation. $p$ value represents the dimension of the document vector generated. But with the increase of $\alpha$, the accuracy levels have dropped slightly, whereas $\alpha = 2$, shows the peak of the 3-dimensional plot.

TABLE I. RESULTS COMPARISON ($doc2vec_{SSM}$, $doc2vec_{NV}$, $doc2vec_{NN}$)

| $\alpha$ Value | | 0.1 | | | 0.5 | | | 1 | | | 2 | | | 3 | | |
|---|---|---|---|---|---|---|---|---|---|---|---|---|---|---|---|---|
| Model | | SSM | NV | NN | SSM | NV | NN | SSM | NV | NN | SSM | NV | NN | SSM | NV | NN |
| $p$ Value | p=250 | 16.26 | **87.70** | 30.05 | 33.89 | **87.70** | 69.70 | 35.01 | **87.70** | 84.04 | 36.04 | **87.70** | 81.13 | 39.21 | 87.70 | **90.34** |
| | p=500 | 23.27 | **87.70** | 41.80 | 39.17 | **87.70** | 85.61 | 39.80 | **87.70** | 88.03 | 39.60 | 87.70 | **89.96** | 42.89 | 87.70 | **90.00** |
| | p=750 | 25.96 | **87.70** | 47.85 | 41.22 | **87.70** | 90.58 | 44.30 | 87.70 | **92.33** | 42.05 | 87.70 | **92.68** | 43.84 | 87.70 | **93.16** |
| | p=1000 | 26.86 | **87.70** | 49.14 | 42.65 | **87.70** | 83.91 | 46.71 | **87.70** | 87.32 | 44.95 | 87.70 | **89.17** | 44.14 | 87.70 | **91.64** |
| | p=2000 | 28.03 | **87.70** | 54.77 | 43.79 | **87.70** | 90.44 | 49.33 | 87.70 | **90.15** | 52.36 | 87.70 | **93.81** | 51.59 | 87.70 | **88.32** |

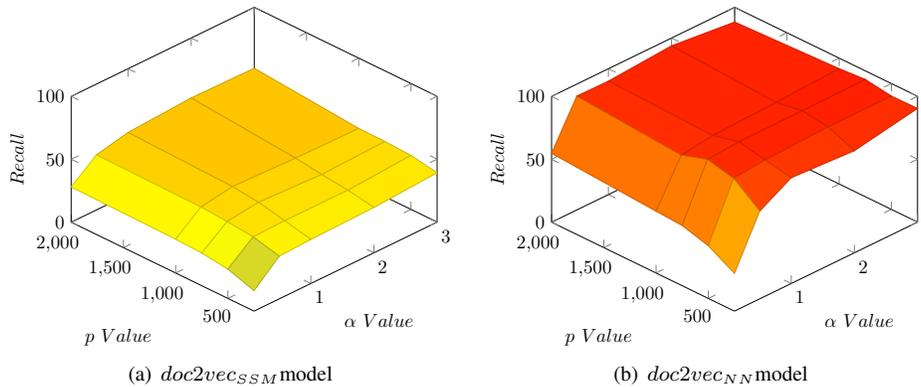

(a) $doc2vec_{SSM}$ model  (b) $doc2vec_{NN}$ model

Fig. 7. 3D plot of accuracy of document vector, against $\alpha$ and p values

## V. CONCLUSION AND FUTURE WORKS

The hypothesis of this study was to prove the high accuracy levels of the $doc2vec_{NN}$ model against $doc2vec_{SSM}$ model and $doc2vec_{NV}$ model. This result is clearly depicted in table I and Fig.7(b). The accuracy levels obtained by the Semantic Processing and Natural Language Processing techniques via the $doc2vec_{SSM}$ model, shows lesser values in terms of results over the $doc2vec_{NV}$ model, which was trained using Node2vec, where input was the Mention Map of this study. But the $doc2vec_{NN}$ model, which is an ensemble of the above two models, gives a significantly higher accuracy level as expected in our hypothesis.

We adopted semantic similarity measures from a previous study and generated a document to vector space to perform document retrieval tasks. This novel approach has shown better accuracy levels, as expected. However, we identified a practical limitation in carrying out this study which we intend to keep as the future work. The number of times a particular legal case is mentioned in a case was not taken into the account in this experiment. It should be noted that, a case which has been mentioned many times has a significant relevance to the case which the case was mentioned in, than a case which has mentioned few times. This could be taken into consideration by allocating a weight to each case based on the number of times it has been mentioned in a case.

The domain of interest in this study was the legal domain, and as mentioned in section I, other domain areas like biology, astronomy, and geology, contain a similar syntactic structure within the domain corpora. The trained model built using our neural network, can be extended towards other domains to create similar information retrieval (IR) systems. Domain specific IR systems stand out in the field due to its complexities and difficulties. Our study has proven the importance of considering domain specific IR systems, which would indeed contribute towards the semantic web development [47].

## REFERENCES

[1] G. Salton and M. Lesk, "Iv information analysis and dictionary construction," 1971.
[2] H.-M. Müller, E. E. Kenny, and P. W. Sternberg, "Textpresso: an ontology-based information retrieval and extraction system for biological literature," *PLoS Biol*, vol. 2, no. 11, p. e309, 2004.
[3] J. Huang, F. Gutierrez, H. J. Strachan *et al.*, "OmniSearch: a semantic search system based on the Ontology for MIcroRNA Target (OMIT) for microRNA-target gene interaction data," *Journal of Biomedical Semantics*, vol. 7, no. 25, 2016.
[4] J. Huang, K. Eilbeck, B. Smith *et al.*, "The development of non-coding RNA ontology," *International journal of data mining and bioinformatics*, vol. 15, no. 3, pp. 214–232, 2016.
[5] D. E. Oliver, Y. Shahar, E. H. Shortliffe, and M. A. Musen, "Representation of change in controlled medical terminologies," *Artificial intelligence in medicine*, vol. 15, no. 1, pp. 53–76, 1999.
[6] W. A. Woods, "Progress in natural language understanding: an application to lunar geology," in *Proceedings of the June 4-8, 1973, national computer conference and exposition*. ACM, 1973, pp. 441–450.
[7] M. Müller, *Information retrieval for music and motion*. Springer, 2007, vol. 2.
[8] C. D. Manning, P. Raghavan, H. Schütze *et al.*, *Introduction to information retrieval*. Cambridge university press Cambridge, 2008, vol. 1, no. 1.
[9] J. Hughes, "Rules for mediation in findlaw for legal professionals," 1999.
[10] G. Salton and C. Buckley, "Term-weighting approaches in automatic text retrieval," *Information processing & management*, vol. 24, no. 5, pp. 513–523, 1988.
[11] J. Sivic, A. Zisserman *et al.*, "Video google: A text retrieval approach to object matching in videos." in *iccv*, vol. 2, no. 1470, 2003, pp. 1470–1477.

[12] D. Greene and P. Cunningham, "Practical solutions to the problem of diagonal dominance in kernel document clustering," in *Proceedings of the 23rd international conference on Machine learning*. ACM, 2006, pp. 377–384.

[13] J. Ramos *et al.*, "Using tf-idf to determine word relevance in document queries," in *Proceedings of the first instructional conference on machine learning*, 2003.

[14] A. Perina, N. Jojic, M. Bicego, and A. Truski, "Documents as multiple overlapping windows into grids of counts," in *Advances in Neural Information Processing Systems*, 2013, pp. 10–18.

[15] J. J. Rocchio, "Relevance feedback in information retrieval," 1971.

[16] G. Salton, A. Singhal, M. Mitra, and C. Buckley, "Automatic text structuring and summarization," *Information Processing & Management*, vol. 33, no. 2, pp. 193–207, 1997.

[17] K. Papineni, S. Roukos, T. Ward, and W.-J. Zhu, "Bleu: a method for automatic evaluation of machine translation," in *Proceedings of the 40th annual meeting on association for computational linguistics*. Association for Computational Linguistics, 2002, pp. 311–318.

[18] V. Jayawardana, D. Lakmal, N. de Silva, A. S. Perera, K. Sugathadasa, B. Ayesha, and M. Perera, "Semi-supervised instance population of an ontology using word vector embeddings," *arXiv preprint arXiv:1709.02911*, 2017.

[19] M. Lapata and R. Barzilay, "Automatic evaluation of text coherence: Models and representations," in *IJCAI*, vol. 5, 2005, pp. 1085–1090.

[20] E. Terra and C. L. Clarke, "Frequency estimates for statistical word similarity measures," in *Proceedings of the 2003 Conference of the North American Chapter of the Association for Computational Linguistics on Human Language Technology-Volume 1*. Association for Computational Linguistics, 2003, pp. 165–172.

[21] K. Sugathadasa, B. Ayesha, N. de Silva, A. S. Perera, V. Jayawardana, D. Lakmal, and M. Perera. (2017) Synergistic union of word2vec and lexicon for domain specific semantic similarity.

[22] R. Mihalcea, C. Corley, C. Strapparava *et al.*, "Corpus-based and knowledge-based measures of text semantic similarity," in *AAAI*, vol. 6, 2006, pp. 775–780.

[23] G. Salton and M. J. McGill, "Introduction to modern information retrieval," 1986.

[24] R. W. Hamming, "Error detecting and error correcting codes," *Bell Labs Technical Journal*, vol. 29, no. 2, pp. 147–160, 1950.

[25] M. Norouzi, D. J. Fleet, and R. R. Salakhutdinov, "Hamming distance metric learning," in *Advances in neural information processing systems*, 2012, pp. 1061–1069.

[26] S.-H. Cha, "Comprehensive survey on distance/similarity measures between probability density functions," *City*, vol. 1, no. 2, p. 1, 2007.

[27] S.-H. Cha, S. Yoon, and C. C. Tappert, "Enhancing binary feature vector similarity measures," 2005.

[28] V. Jayawardana, D. Lakmal, N. de Silva, A. S. Perera, K. Sugathadasa, and B. Ayesha, "Deriving a representative vector for ontology classes with instance word vector embeddings," *arXiv preprint arXiv:1706.02909*, 2017.

[29] Z. Wu and M. Palmer, "Verbs semantics and lexical selection," in *Proceedings of the 32Nd Annual Meeting on Association for Computational Linguistics*, ser. ACL '94. Stroudsburg, PA, USA: Association for Computational Linguistics, 1994, pp. 133–138. [Online]. Available: http://dx.doi.org/10.3115/981732.981751

[30] G. Qian, S. Sural, Y. Gu, and S. Pramanik, "Similarity between euclidean and cosine angle distance for nearest neighbor queries," in *Proceedings of the 2004 ACM symposium on Applied computing*. ACM, 2004, pp. 1232–1237.

[31] T. Mikolov, K. Chen, G. Corrado, and J. Dean, "Efficient estimation of word representations in vector space," *arXiv preprint arXiv:1301.3781*, 2013.

[32] D. A. Evans, J. J. Cimino, W. R. Hersh, S. M. Huff, D. S. Bell *et al.*, "Toward a medical-concept representation language," *Journal of the American Medical Informatics Association*, vol. 1, no. 3, pp. 207–217, 1994.

[33] B. Tang, H. Cao, Y. Wu, M. Jiang, and H. Xu, "Recognizing clinical entities in hospital discharge summaries using structural support vector machines with word representation features," *BMC medical informatics and decision making*, vol. 13, no. 1, p. S1, 2013.

[34] E. Schweighofer and W. Winiwarter, "Legal expert system kontermautomatic representation of document structure and contents," in *International Conference on Database and Expert Systems Applications*. Springer, 1993, pp. 486–497.

[35] J. J. Nay, "Gov2vec: Learning distributed representations of institutions and their legal text," *arXiv preprint arXiv:1609.06616*, 2016.

[36] Y. Bengio, R. Ducharme, P. Vincent, and C. Jauvin, "A neural probabilistic language model," *Journal of machine learning research*, vol. 3, no. Feb, pp. 1137–1155, 2003.

[37] R. Mihalcea and P. Tarau, "Textrank: Bringing order into texts." Association for Computational Linguistics, 2004.

[38] L. Page, S. Brin, R. Motwani, and T. Winograd, "The pagerank citation ranking: Bringing order to the web." Stanford InfoLab, Tech. Rep., 1999.

[39] D. R. Radev, "A common theory of information fusion from multiple text sources step one: cross-document structure," in *Proceedings of the 1st SIGdial workshop on Discourse and dialogue-Volume 10*. Association for Computational Linguistics, 2000, pp. 74–83.

[40] Q. Le and T. Mikolov, "Distributed representations of sentences and documents," in *Proceedings of the 31st International Conference on Machine Learning (ICML-14)*, 2014, pp. 1188–1196.

[41] Y. Zhang, R. Jin, and Z.-H. Zhou, "Understanding bag-of-words model: a statistical framework," *International Journal of Machine Learning and Cybernetics*, vol. 1, no. 1-4, pp. 43–52, 2010.

[42] C. D. Manning, M. Surdeanu, J. Bauer, J. R. Finkel, S. Bethard, and D. McClosky, "The stanford corenlp natural language processing toolkit." in *ACL (System Demonstrations)*, 2014, pp. 55–60.

[43] X. Wang, L. Wang, and Y. Qiao, "A comparative study of encoding, pooling and normalization methods for action recognition," in *Asian Conference on Computer Vision*. Springer, 2012, pp. 572–585.

[44] A. Grover and J. Leskovec, "node2vec: Scalable feature learning for networks," in *Proceedings of the 22nd ACM SIGKDD international conference on Knowledge discov-

Computing Conference 2018
10-12 July 2018 — London, UK*ery and data mining*. ACM, 2016, pp. 855–864.

[45] D. C. Wimalasuriya and D. Dou, "Ontology-based information extraction: An introduction and a survey of current approaches," *Journal of Information Science*, 2010.

[46] R. Kohavi *et al.*, "A study of cross-validation and bootstrap for accuracy estimation and model selection," in *Ijcai*, vol. 14, no. 2. Stanford, CA, 1995, pp. 1137–1145.

[47] T. Berners-Lee, J. Hendler, O. Lassila *et al.*, "The semantic web," *Scientific american*, vol. 284, no. 5, pp. 28–37, 2001.9 | Page